\documentclass{article}
\usepackage{spconf,amsmath,graphicx}
\usepackage{amsfonts}
\usepackage{epsfig,amssymb,latexsym,graphics,float}
\usepackage{setspace,subfig}
\usepackage{citesort}
\usepackage{amsopn}
\usepackage{amssymb,times}
\usepackage{ifpdf}
\usepackage{algorithm}
\usepackage{algorithmic}

\title{Data-Driven Model Set Design for Model Averaged Particle Filter}
\name{Bin~Liu$^{\ast}$
\thanks{$^\ast$Address correspondence to bins@ieee.org. This work was partly supported by National key research and development plan of China (No. YFB2101704), and National Natural Science Foundation (NSF) of
China (Nos. 61571238 and 61906099).}}
\address{School of Computer Science, \\
Jiangsu Key Lab of Big Data Security $\&$ Intelligent Processing, \\
Nanjing University of Posts and Telecommunications, Nanjing, 210023 China}
\begin{document}
\maketitle
\begin{abstract}
This paper is concerned with sequential state filtering in the presence of nonlinearity, non-Gaussianity and model uncertainty. For this problem, the Bayesian model averaged particle filter (BMAPF) is perhaps one of the most efficient solutions. Major advances of BMAPF have been made, while it still lacks a generic and practical approach to design the model set. This paper fills in this gap by proposing a generic data-driven method for BMAPF model set design. Unlike existent methods, the proposed solution does not require any prior knowledge on the parameter value of the true model; it only assumes that a small number of noisy observations are pre-obtained. The Bayesian optimization (BO) method is adapted to search the model components, each of which is associated with a specific segment of the pre-obtained dataset.
The average performance of these model components is guaranteed since each one's parameter is elaborately tuned via BO to maximize the marginal likelihood. The diversity in the model components is also ensured, as different components match the different segments of the pre-obtained dataset, respectively. Computer simulations are used to demonstrate the effectiveness of the proposed method.
\end{abstract}
\section{Introduction}\label{sec:intro}
This paper is concerned with the state-space model based sequential state filtering, which finds widespread applications in signal processing, statistics, and econometrics. In particular, we focus on cases in which the state-space model can be nonlinear, non-Gaussian as well as uncertain. Since \cite{gordon1993novel}, particle filtering (PF) has become the dominated methodology that addresses problems cast by nonlinear non-Gaussian state-space models \cite{arulampalam2002tutorial,gordon1993novel}. Most of PF based sequential filtering methods assume that the underlying model is deterministic, with few notable exceptions in e.g, \cite{nemeth2013sequential,liu2001combined,kantas2009overview}. Once the model being used mismatches the true one, PF shall almost certainly fail. To mitigate the risk of model mismatch, one has to take into account model uncertainty in the design of a sequential filtering algorithm. A common strategy is to employ a set of different models. The combination of such multi-model strategy and PF yields the multi-model based PF (MMPF) methods \cite{boers2003interacting,urteaga2016sequential,drovandi2014sequential,martino2017cooperative}.

In our previous work, we proposed a specific type of MMPF, termed Bayesian model averaged PF (BMAPF) here. A basic feature of BMAPF lies in the application of the Bayesian model averaging theory for adjusting the effect of each model component online. In BMAPF, the model components are weighted according to their posterior probabilities rather than heuristics. The posterior probability of each model is approximately calculated and updated in virtue of the weighed particle set of PF. The BMAPF methodology has been successfully used in different applications. In \cite{liu2011instantaneous}, we use BMAPF for tracking the instantaneous frequency of a non-stationary signal.
%wherein the temporal evolution law of the IF is captured by multiple models.
In \cite{yi2016robust}, we make use of BMAPF for video object feature fusion and robust tracking. In \cite{liu2017robust}, we present a sequential filtering algorithm based on BMAPF, which is robust against the presence of outliers in the measurements. In our most recent work in \cite{qi2019dynamic}, BMAPF is adapted to deal with non-stationary neural decoding in Brain-computer interfaces, and its performance has been demonstrated using real neural datasets. Despite advances that have been achieved, one essential concern on practical applications of BMAPF remains, namely:
\begin{itemize}
\item When there is not adequate prior knowledge for use, how to design a qualified model set for BMAPF?
\end{itemize}

The above concern motivates this work. To address it, we resort to the literature on ensemble neural networks (ENN), for which the issue of model set design (MSD) has been investigated. It is shown that, in the context of ENN, the average performance and the diversity of the model components constitute two essential factors for a successful MSD \cite{krogh1995neural,brown2003use,zhou2002ensembling}. We investigate here whether they are also crucial factors for MSD of BMAPF. Specifically, we develop an MSD assisted BMAPF (MSD-BMAPF), in which the average performance and the diversity of the model components are balanced. We demonstrate that MSD-BMAPF outperforms the baseline BMAPF with simulated experiments.

To summarize, the main contribution of this paper is twofold. First, we confirm that the average performance and the diversity of the model components are important factors for MSD of BMAPF. Second, we propose a novel algorithm, MSD-BMAPF, which targets for cases when one lacks prior knowledge for specifying a qualified model set for BMAPF.
\section{Preliminary}
\subsection{A Baseline BMAPF Algorithm}\label{sec:bmapf}
Consider a state-space model $\mathcal{M}$ defined by a state-transition prior density function $p_{\theta}(\textbf{x}_t|\textbf{x}_{t-1})$ and a likelihood function $p_{\theta}(\textbf{y}_t|\textbf{x}_t)$, where $t$ denotes the discrete-time index, $\textbf{x}\in\mathbb{R}^{d_x}$ the state of interest to be estimated, $\textbf{y}\in\mathbb{R}^{d_y}$ the measurement observed, and $\theta\in\Theta$ denotes the model parameter whose value is not \emph{a priori} known.

The task is, at each time step $t$, to infer the hidden state $\textbf{x}_t$ given measurements that have been observed so far, namely $\textbf{y}_{1:t}\triangleq[\textbf{y}_1,\ldots,\textbf{y}_t]$.
To handle model uncertainty resulted from the unknown value of $\theta$, BMAPF makes uses of multiple models $\mathcal{M}_1,\ldots,\mathcal{M}_K$. Each $\mathcal{M}_k$ is assigned with a specific parameter value of $\theta$, denoted by $\theta_k\in\Theta$, $k=1,\ldots,K$. Let $\mathcal{H}_t=k$ denote the hypothesis that $\mathcal{M}_k$ is the true model of the system at time $t$. Based on the Bayesian model averaging theory \cite{hoeting1999bayesian,raftery1997bayesian}, the posterior probabilistic density function (pdf) of $\textbf{x}_t$, denoted by $p_{t|t}\triangleq p(\textbf{x}_t|\textbf{y}_{1:t})$, can be calculated as follows
\begin{equation}\label{eqn:bma}
p_{t|t}=\sum_{k=1}^Kp_{k,t|t}\pi_{k,t},
\end{equation}
where $p_{k,t|t}\triangleq p(\textbf{x}_t|\mathcal{H}_t=k,\textbf{y}_{1:t})$ and $\pi_{k,t}\triangleq p(\mathcal{H}_{t}=k|\textbf{y}_{1:t})$, for $t\geq1$.

The BMAPF provides a recursive solution to compute Eqn.(\ref{eqn:bma}).
It is initialized by specifying a prior density $p(\textbf{x}_0)$ of $\textbf{x}_0$, and defining $p_{0|0}=p(\textbf{x}_0)$.
At time $t$, $t\geq1$, a parallel of $K$ importance sampling (IS) procedures are run, each corresponding to a model component under consideration. In the $k$th IS procedure, first, draw $\textbf{x}_{k,t-1}^i$ from $p_{t-1|t-1}, i=1,\ldots,N_k$,
%where $\sim$ means ``is distributed according to",
where $N_k$ denotes the number of particles associated with $\mathcal{M}_k$, $k=1,\ldots,K$. The associated particle weights are $\omega_{k,t-1}^i=1/N_k, i=1,\ldots,N_k$.
A Monte Carlo approximation of $p_{t-1|t-1}$ is
\begin{equation}\label{eqn:particle_approx_t-1}
p_{t-1|t-1}\simeq\sum_{i=1}^{N_k}\left[\omega_{k,t-1}^i\delta(\textbf{x}_{t-1},\textbf{x}_{k,t-1}^i)\right],
\end{equation}
where the Kronecker-delta function $\delta(a,b) = 1$ if and only if $a = b$ and 0 otherwise.
Draw $\hat{\textbf{x}}_{k,t}^i$ from $p_{\theta_k}(\textbf{x}_t|\textbf{x}_{k,t-1}^i), i=1,\ldots,N_k$, then it leads to
\begin{equation}
p_{k,t|t-1}\simeq\sum_{i=1}^{N_k}\left[\omega_{k,t-1}^i\delta(\textbf{x}_t,\hat{\textbf{x}}_{k,t}^i)\right],
\end{equation}
where $p_{k,t|t-1}\triangleq p(\textbf{x}_t|\mathcal{H}_t=k,\textbf{y}_{1:t-1})$. The marginal likelihood or evidence of the observation $\textbf{y}_t$ under the hypothesis $\mathcal{H}_t=k$ is $L_{k,t}\triangleq\int_{\mathcal{X}}p_{\theta_k}(\textbf{y}_t|\textbf{x}_t)p_{k,t|t-1}d\textbf{x}_t$, which can be unbiasedly approximated as follows \cite{moral2004feynman}
\begin{equation}\label{eq:marglik_approx}
L_{k,t}\thickapprox \sum_{i=1}^{N_k}\hat{\omega}_{k,t}^i, \hat{\omega}_{k,t}^i=\omega_{k,t-1}^i p_{\theta_k}(\textbf{y}_t|\hat{\textbf{x}}_{k,t}^i), i=1,\ldots,N_k.
\end{equation}

According to the theory of IS (see details in \cite{arulampalam2002tutorial}), if one assigns weight $\omega_{k,t}^i$ to $\hat{x}_{k,t}^i$, where
\begin{equation}\label{eqn:omega_mm}
\omega_{k,t}^i=\frac{\hat{\omega}_{k,t}^i}{\sum_{j=1}^{N_k}\hat{\omega}_{k,t}^j}, i=1,\ldots,N_k,
\end{equation}
then it leads to
\begin{equation}\label{eqn:particle_approx}
p_{k,t|t}\simeq\sum_{i=1}^{N_k}\left[\omega_{k,t}^i\delta(\textbf{x}_t,\hat{\textbf{x}}_{k,t}^i)\right].
\end{equation}

Let $\pi_{k,t-1}$ act as the prior probability of the hypothesis $\mathcal{H}_t=k$, $k=1,\ldots,K$, then, the Bayes theorem tells that
\begin{equation}\label{eq:post_prob_model}
\pi_{k,t}=\frac{\pi_{k,t-1}L_{k,t}}{\sum_{j=1}^K\pi_{j,t-1}L_{j,t}}, k=1,\ldots,K.
\end{equation}
Substituting Eqns. (\ref{eqn:particle_approx}) and (\ref{eq:post_prob_model}) into Eqn. (\ref{eqn:bma}), one then obtains a particle-based approximation to $p_{t|t}$ as follows
\begin{equation}\label{eq:particle_approx_posterior}
p_{t|t}\thickapprox \sum_{k=1}^K\left[\pi_{k,t}\sum_{i=1}^{N_k}\left[\omega_{k,t}^i\delta(\textbf{x}_t,\hat{\textbf{x}}_{k,t}^i)\right]\right].
\end{equation}
Any statistics, e.g., mean, variance, about $\textbf{x}_t$ can be obtained based on the weighted particle set
$\{\{\hat{\textbf{x}}_{k,t}^i, \pi_{k,t}\omega_{k,t}^i\}_{i=1}^{N_k}\}_{k=1}^K$.
\begin{algorithm}[!htb]
\caption{The baseline BMAPF Algorithm}
\label{alg:BMAPF}
\begin{algorithmic}[1]
\STATE Initialization: Specify $\theta_1,\ldots,\theta_K$ by domain knowledge. Let $\pi_{k,0}=1/K, k=1,\ldots,K$; Specify $p(\textbf{x}_0)$, and let $p_{0|0}=p(\textbf{x}_0)$.
\FOR{$t$=1,2,\ldots}
\FOR{$k$=1,\ldots,K}
\STATE Resampling: draw $\textbf{x}_{k,t-1}^i$ from $p_{t-1|t-1}$ and set $\omega_{k,t-1}^i=1/N_k, i=1,\ldots,N_k$.
\STATE Propagation: draw $\hat{\textbf{x}}_{k,t}^i$ from $p_{\theta_k}(\textbf{x}_t|\textbf{x}_{k,t-1}^i), i=1,\ldots,N_k$.
\STATE Calculate $L_{k,t}$ with Eqn. (\ref{eq:marglik_approx}).
\STATE Calculate $\omega_{k,t}^i, i=1,\ldots,N_k$, with Eqn. (\ref{eqn:omega_mm}).
\ENDFOR
\STATE Calculate $\pi_{k,t}$ with Eqn. (\ref{eq:post_prob_model}), $k=1,\ldots,K$.
\STATE Output: weighted particle set $\{\{\hat{\textbf{x}}_{k,t}^i, \pi_{k,t}\omega_{k,t}^i\}_{i=1}^{N_k}\}_{k=1}^K$ and an approximation of $p_{t|t}$, see Eqn. (\ref{eq:particle_approx_posterior}).
\ENDFOR
\end{algorithmic}
\end{algorithm}

A summarization of this baseline algorithm is presented in Algorithm \ref{alg:BMAPF}. Note that the operation at the 4th line of Algorithm \ref{alg:BMAPF}, i.e., drawing samples $\textbf{x}_{k,t-1}^i$ from $p_{t-1|t-1}$, is actually a resampling operation, since only an approximated discrete measure of $p_{t-1|t-1}$, yielded at the previous iteration, is available. Such a resampling operation is necessary for getting rid of the issue of particle degeneracy \cite{arulampalam2002tutorial}. For brevity, $N_k$ is treated as a pre-set constant here, while it can also be adapted online as proposed in \cite{martino2017cooperative}.
%Such adaptive tuning approach is adopted in our experiments as presented in Section \ref{sec:experiment}.
\subsection{Bayesian optimization (BO)}\label{sec:BO}
Consider a maximization problem $\underset{\theta\in\Theta}{\max}f(\theta)$, where $f$: $\Theta\rightarrow\mathbb{R}$ is a smooth real-valued function. The global optimum is denoted by $\theta^{\ast}\in\Theta$, namely, $f(\theta)\leq f(\theta^{\ast})$, $\forall \theta\in\Theta$.
BO through Gaussian Process (GP) regression is an efficient methodology for searching $\theta^{\ast}$, especially when no further assumptions on $f$ can be made. The basic idea is to estimate the distribution over function $f$ via the GP nonparametric approach, and then use this information to decide the next point of $f$ to be evaluated.

In the GP approach, the prior of $f$ is described by a GP process, $GP(\mu(\cdot),\kappa(\cdot))$, where $\mu(\cdot)$ is the mean function and $\kappa(\cdot)$ is the covariance function. That says $\mu(\theta)=\mathbb{E}[f(\theta)]$ and $\kappa(\theta,\theta')=\mathbb{E}[(f(\theta)-\mu(\theta))(f(\theta')-\mu(\theta'))^T]$, where $A^T$ denotes the transposition of $A$.
Suppose that $n$ evaluations of $f$ have been made, denoted by $\mathcal{D}_n:=\{\theta_{1:n},\mathbf{f}(\theta_{1:n})\}$, where $\mathbf{f}(\theta_{1:n}):=[f(\theta_1),\ldots,f(\theta_n)]$. Under the above setting, we have $\mathbf{f}(\theta_{1:n})\sim\mathcal{N}(\mu(\theta_{1:n}),\mathbf{K})$, where $\mathbf{K}_{i,j}=\kappa(\theta_i,\theta_j)$.
Given a new data point $\theta_{n+1}$, the joint distribution over $\mathbf{f}(\theta_{1:n})$ and $f(\theta_{n+1})$ is also Gaussian:
\begin{equation}
\left(\begin{array}{c}{\mathbf{f}\left(\mathbf{\theta}_{1:n}\right)} \\ {f\left(\theta_{n+1}\right)}\end{array}\right) \sim \mathcal{N}\left(\begin{array}{c}{\mathbf{\mu}\left(\theta_{1:n}\right)} \\ {\mu\left(\theta_{n+1}\right)}\end{array},\left[\begin{array}{cc}{\mathbf{K}} & {\mathbf{k}} \\ {\mathbf{k}^{T}} & {\kappa\left(\theta_{n+1}, \theta_{n+1}\right)}\end{array}\right]\right)
\end{equation}
where $\mathbf{k}=\kappa\left(\theta_{1:n},\theta_{n+1}\right) \in \mathbb{R}^{n\times1}$. Then the predictive distribution of $f(\theta_{n+1})$ conditioned on $\mathcal{D}_n$ and $\theta_{n+1}$ is:
\begin{equation}
f\left(\theta_{n+1}\right)\left|\mathcal{D}_{n}, \theta_{n+1} \sim \mathcal{N}\left(\mu\left(\theta_{n+1}| \mathcal{D}_{n}\right), \sigma^{2}\left(\theta_{n+1}| \mathcal{D}_{n}\right)\right)\right.
\end{equation}
where $\mu\left(\theta_{n+1}|\mathcal{D}_{n}\right)=\mu\left(\theta_{n+1}\right)+\mathbf{k}^{T} \mathbf{K}^{-1}\left(\mathbf{f}\left(\theta_{1:n}\right)-\mathbf{\mu}\left(\theta_{1:n}\right)\right)$
and $\sigma^{2}\left(\theta_{n+1}| \mathcal{D}_{n}\right)=\kappa\left(\theta_{n+1}, \theta_{n+1}\right)-\mathbf{k}^{T} \mathbf{K}^{-1} \mathbf{k}$.
Usually $\mu(\cdot)$ is set to be a zero-valued function. Common choices of covariance functions include the Matern kernel and the Gaussian
kernel. The Gaussian kernel is defined to be $\kappa\left(\theta, \theta^{\prime}\right)=\exp \left(-\frac{1}{2}\left(\theta-\theta^{\prime}\right)^{T} \Sigma^{-1}\left(\theta-\theta^{\prime}\right)\right)$, where $\Sigma^{-1}$ is the kernel parameter matrix, which can be learned from data or specified by the user based on domain knowledge. For more details about GP, see \cite{williams2006gaussian}.

Given observed data points $\mathcal{D}_n$, BO selects the next query point that optimizes an acquisition function generated by GP, such as upper confidence bound (UCB) and expected improvement (EI). BO with GP-UCB sets $\theta_{n+1}=\arg\max_{\theta}[\mu\left(\theta|\mathcal{D}_{n}\right)+\alpha\sigma\left(\theta|\mathcal{D}_{n}\right)]$, where $\alpha\in\mathbb{R}$ is a parameter of the algorithm that tradeoffs exploration and exploitation. A BO procedure implemented by means of GP-UCB is shown in Algorithm \ref{alg:GP-UCB}. See \cite{shahriari2016taking} for more details about BO.
\begin{algorithm}[tb]
\caption{A BO procedure based on GP-UCB}
\label{alg:GP-UCB}
\begin{algorithmic}[1] %[1] enables line numbers
\FOR{$n$=1,2,\ldots}
\STATE Select $\theta_{n+1}$ such that $\theta_{n+1}=\arg\max_{\theta}[\mu\left(\theta|\mathcal{D}_{n}\right)+\alpha\sigma\left(\theta|\mathcal{D}_{n}\right)]$;
\STATE Augment the observed data set by setting $\mathcal{D}_{n+1}:=\{\theta_{1:n+1},\mathbf{f}(\theta_{1:n+1})\}$.
\STATE If the loop termination condition is satisfied, take out of the loop.
\ENDFOR
\STATE Return $\theta^{\ast}=\arg\max_{\theta_i\in\theta_{1:n+1}}f(\theta_i)$.
\end{algorithmic}
\end{algorithm}
%This additional
% non-convex optimization is often carried out by other global optimization methods such as DIRECT and CMA-ES.
\section{The Proposed BO based MSD Approach}\label{sec:BO_MSD}
\begin{figure}[!htb]
\centering
\includegraphics[width=3in,height=1.7in]{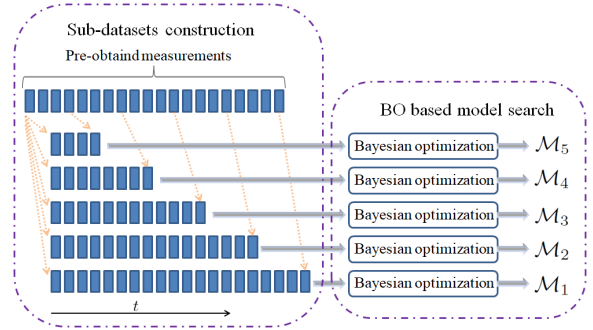}
\caption{An example illustration of the proposed BO based MSD approach. In this example, $K=5$ model components are built based on 20 data blocks of pre-obtained observations.}\label{fig:MSD-BMAPF}
\end{figure}

Suppose that one has a set of pre-obtained measurements, $\textbf{y}_{1:m}, m\in\mathbb{R}$, while has no prior knowledge to specify a model set for use in BMAPF. We now present a data-driven approach to building up a set of candidate models based on $\textbf{y}_{1:m}$. We propose a BO based MSD (BOMSD) approach, as shown in Fig.\ref{fig:MSD-BMAPF}, which makes a balance between the averaged performance and the diversity of the model components.

Specifically, first, construct a series of sub-datasets, $\textbf{y}_{1:m_k}$, $m_k=\lfloor m(K-k+1)/K\rfloor$, $k=1,\ldots,K$, where $\lfloor a\rfloor$ denotes the biggest integer that does not exceed $a$.
Then, for each sub-dataset, say $\textbf{y}_{1:m_k}$, run a BO algorithm to search one model component, $\mathcal{M}_k$, whose parameter value $\theta_k$ maximizes the objective function $f_k(\theta)\triangleq\log(p_\theta(\textbf{y}_{1:m_k}))=\sum_{t=1}^{m_k}\log(L_{k,t}(\theta))$. Here $L_{k,t}(\theta)$ is the evidence of $\textbf{y}_t$ associated with $\theta$, namely,
\begin{equation}\label{eq:marglik3}
L_{k,t}(\theta)\triangleq\int_{\mathcal{X}}p_{\theta}(\textbf{y}_t|\textbf{x}_t)p_{k,t|t-1}d\textbf{x}_t.
\end{equation}
As shown in subsection \ref{sec:bmapf}, given a $\theta$ value, one can approximate $L_{k,t}(\theta)$ in virtue of the weighted particle set of a PF, see Eqn.(\ref{eq:marglik_approx}). That says, for any query point $\theta$, we could run a PF to yield an evaluation of it. It indicates that the objective function $f_k$ is expensive to evaluate. Therefore we adopt BO here for function optimization.
\section{Experiments}\label{sec:experiment}
We did two experiments to test whether the proposed BOMSD approach is effective, by comparing MSD-BMAPF with the baseline BMAPF. The only difference between MSD-BMAPF and the baseline BMAPF lies in that, for the former, the model components' parameter values are initialized by our BOMSD approach; for the latter, they are randomly initialized. We adopted a state-of-the-art BO method \cite{kawaguchi2015bayesian}, which has an exponential convergence rate, to implement MSD-BMAPF, ensuring the efficiency of model search.
\subsection{Experiment I}
\begin{figure}[!htb]
\centering
\includegraphics[width=3in,height=2in]{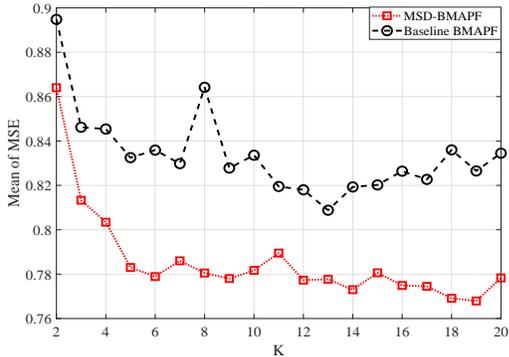}
\caption{Performance comparison between MSD-BMAPF and the baseline BMAPF in Experiment I. $K$ denotes the number of model components employed by the algorithms.  For each $K$ value between 2 and 20, the averaged MSE over 100 independent runs of each algorithm is plotted.}\label{fig:mse}
\end{figure}

In the first experiment, the state-space model under consideration is
\begin{equation}
\left\{\begin{array}{l}{\textbf{x}_{t}=\theta\left|\textbf{x}_{t-1}\right|+\textbf{v}_{t}} \\
{\textbf{y}_{t}=\log \left(\textbf{x}_{t}^{2}\right)+\textbf{u}_{t}}\end{array}, \quad t=1, \ldots, 500\right.
\end{equation}
where $\textbf{x}_0=0$, $0\leq\theta\leq1$, the value of $\theta$ is not \emph{a priori} known, $\textbf{v}_{t}\sim\mathcal{N}(0,1)$, and $\textbf{u}_{t}\sim\mathcal{N}(0,1)$. The parameter value of the true model that generates the observations is $\theta^{\ast}=0.657$. In the baseline BMAPF, the $\theta$ value of each model component is randomly chosen from between 0 and 1. MSD-BMAPF searches the model components via BO based on 200 historical data points generated by the same model. We considered 19 cases, in which $K=2,\ldots,20$, respectively. For each case, 100 independent runs of each algorithm are performed. Then we computed the averaged mean squared error (MSE) over these 100 runs and plot the result in Fig.\ref{fig:mse}. As is shown, for every $K$ value considered, our MSD-BMAPF algorithm outperforms the baseline BMAPF markedly. It also shows that, when the number of models $K$ reaches a threshold, employing more models does not necessarily lead to better performance.
\subsection{Experiment II}
\begin{figure}[!htb]
\centering
\includegraphics[width=3in,height=2in]{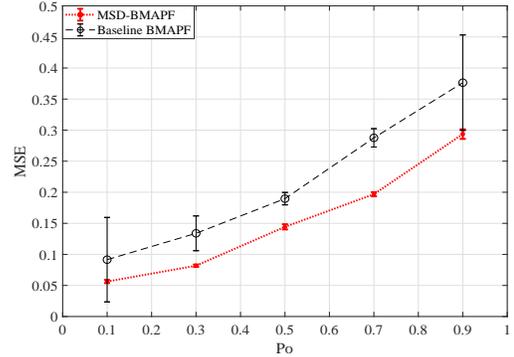}
\caption{Results of MSD-BMAPF and the baseline BMAPF in Experiment II. It shows two standard deviations and the mean of the MSE for 5 cases, in which the true $P_o$ takes values at 0.1, 0.3, 0.5, 0.7 and 0.9, respectively.}\label{fig:error_bar}
\end{figure}

The second experiment is borrowed from \cite{liu2019robust}. The state transition function is
\begin{equation}
\textbf{x}_{t+1}=1+\sin\left(\frac{4\pi\mbox{mod}(t+1,60)}{100}\right)+0.5\textbf{x}_t+\textbf{u}_t,
\end{equation}
where $1\leq t<600$, $\textbf{x}_1=1$, $\textbf{u}_t\sim Gamma(3,2)$, mod($a,b$) denotes the remainder after the division of $a$ by $b$. The measurement function is
\begin{equation}\label{measure_func_simu}
\textbf{y}_t=\left\{\begin{array}{ll}
0.2\textbf{x}_t^2+\textbf{n}_t,\quad\quad\; \mbox{if}\;\mbox{mod}(t,60)\leq30 \\
0.2\textbf{x}_t-2+\textbf{n}_t,\quad \mbox{otherwise} \end{array} \right.
\end{equation}
where $\textbf{n}_t$ is with probability $P_o$ distributed according to $0.5\mathcal{N}(\cdot|20,0.1)+0.5\mathcal{N}(\cdot|22,0.1)$, and with probability $1-P_o$ to $\mathcal{N}(\cdot|0,0.01)$, $0\leq P_o\leq1$. In this experiment, $K$ is fixed at 3 for both methods and the uncertainty of the model is reflected by the unknown value of $P_o$. We considered five cases, in which the true $P_o$ takes values at $0.1, 0.3, 0.5, 0.7, 0.9$, respectively. Again, for each case, we run both algorithms 100 times. For each algorithm under each case, we calculated the mean and the standard error of its MSE. The result is plotted in Fig.\ref{fig:error_bar}. MSD-BMAPF again performs significantly better than the baseline BMAPF.
\section{Concluding Remarks}
BMAPF finds widespread and successful applications for sequential filtering under model uncertainty. Here we presented a
generic yet practical method, namely BOMSD, for automatic
MSD of BMAPF. We demonstrated a remarkable performance of our method via simulated experiments. In the future, it is interesting to compare BMAPF with other related methods, such as the Liu and West filter \cite{liu2001combined}, and to investigate how to adapt BMAPF to handle more complex scenarios wherein the structure of the base model is not \emph{a priori} known.
\newpage
\bibliographystyle{IEEEbib}
\bibliography{mybibfile}

\begin{thebibliography}{10}

\bibitem{gordon1993novel}
N.~Gordon, D.~Salmond, and A.~F.~M. Smith,
\newblock ``Novel approach to nonlinear/non-{G}aussian {B}ayesian state
  estimation,''
\newblock {\em IEE Proceedings F (Radar and Signal Processing)}, vol. 140, no.
  2, pp. 107--113, 1993.

\bibitem{arulampalam2002tutorial}
M.~S. Arulampalam, S.~Maskell, N.~Gordon, and T.~Clapp,
\newblock ``A tutorial on particle filters for online nonlinear/non-{G}aussian
  {B}ayesian tracking,''
\newblock {\em IEEE Trans. on Signal Processing}, vol. 50, no. 2, pp. 174--188,
  2002.

\bibitem{nemeth2013sequential}
C.~Nemeth, P.~Fearnhead, and L.~Mihaylova,
\newblock ``Sequential {M}onte {C}arlo methods for state and parameter
  estimation in abruptly changing environments,''
\newblock {\em IEEE Trans. on Signal Processing}, vol. 62, no. 5, pp.
  1245--1255, 2013.

\bibitem{liu2001combined}
J.~Liu and M.~West,
\newblock ``Combined parameter and state estimation in simulation-based
  filtering,''
\newblock in {\em Sequential {M}onte {C}arlo methods in practice}, pp.
  197--223. Springer, 2001.

\bibitem{kantas2009overview}
N.~Kantas, A.~Doucet, S.~S. Singh, and J.~M. Maciejowski,
\newblock ``An overview of sequential {M}onte {C}arlo methods for parameter
  estimation in general state-space models,''
\newblock {\em IFAC Proceedings Volumes}, vol. 42, no. 10, pp. 774--785, 2009.

\bibitem{boers2003interacting}
Y.~Boers and J.~N. Driessen,
\newblock ``Interacting multiple model particle filter,''
\newblock {\em IEE Proceedings-Radar, Sonar and Navigation}, vol. 150, no. 5,
  pp. 344--349, 2003.

\bibitem{urteaga2016sequential}
I.~Urteaga, M.~F. Bugallo, and P.~M. Djuri{\'c},
\newblock ``Sequential {M}onte {C}arlo methods under model uncertainty,''
\newblock in {\em 2016 IEEE Statistical Signal Processing Workshop (SSP)}.
  IEEE, 2016, pp. 1--5.

\bibitem{drovandi2014sequential}
C.~C. Drovandi, J.~M. McGree, and A.~N. Pettitt,
\newblock ``A sequential {M}onte {C}arlo algorithm to incorporate model
  uncertainty in {B}ayesian sequential design,''
\newblock {\em Journal of Computational and Graphical Statistics}, vol. 23, no.
  1, pp. 3--24, 2014.

\bibitem{martino2017cooperative}
L.~Martino, J.~Read, V.~Elvira, and F.~Louzada,
\newblock ``Cooperative parallel particle filters for online model selection
  and applications to urban mobility,''
\newblock {\em Digital Signal Processing}, vol. 60, pp. 172--185, 2017.

\bibitem{liu2011instantaneous}
B.~Liu,
\newblock ``Instantaneous frequency tracking under model uncertainty via
  dynamic model averaging and particle filtering,''
\newblock {\em IEEE Trans. on Wireless Communications}, vol. 10, no. 6, pp.
  1810--1819, 2011.

\bibitem{yi2016robust}
Y.~Dai and B.~Liu,
\newblock ``Robust video object tracking via {B}ayesian model averaging-based
  feature fusion,''
\newblock {\em Optical Engineering}, vol. 55, no. 8, pp. 083102(1)--083102(11),
  2016.

\bibitem{liu2017robust}
B.~Liu,
\newblock ``Robust particle filter by dynamic averaging of multiple noise
  models,''
\newblock in {\em Proc. of the 42nd IEEE Int'l Conf. on Acoustics, Speech, and
  Signal Processing (ICASSP)}. IEEE, 2017, pp. 4034--4038.

\bibitem{qi2019dynamic}
Y.~Qi, B.~Liu, Y.~Wang, and G.~Pan,
\newblock ``Dynamic ensemble modeling approach to nonstationary neural decoding
  in {B}rain-computer interfaces,''
\newblock in {\em Advances in neural information processing systems (NeurIPS)},
  2019, pp. 6087--6096.

\bibitem{krogh1995neural}
A.~Krogh and J.~Vedelsby,
\newblock ``Neural network ensembles, cross validation, and active learning,''
\newblock in {\em Advances in neural information processing systems (NIPS)},
  1995, pp. 231--238.

\bibitem{brown2003use}
G.~Brown and J.~L. Wyatt,
\newblock ``The use of the ambiguity decomposition in neural network ensemble
  learning methods,''
\newblock in {\em Proc. of the 20th Int'l Conf. on Machine Learning (ICML)},
  2003, pp. 67--74.

\bibitem{zhou2002ensembling}
Z.~Zhou, J.~Wu, and W.~Tang,
\newblock ``Ensembling neural networks: many could be better than all,''
\newblock {\em Artificial intelligence}, vol. 137, pp. 239--263, 2002.

\bibitem{hoeting1999bayesian}
J.A. Hoeting, D.~Madigan, A.E. Raftery, and C.T. Volinsky,
\newblock ``{Bayesian model averaging: A tutorial},''
\newblock {\em Statistical science}, vol. 14, no. 4, pp. 382--401, 1999.

\bibitem{raftery1997bayesian}
A.E. Raftery, D.~Madigan, and J.A. Hoeting,
\newblock ``{Bayesian model averaging for linear regression models},''
\newblock {\em Journal of the American Statistical Association}, vol. 92, no.
  437, pp. 179--191, 1997.

\bibitem{moral2004feynman}
P.~Del~Moral,
\newblock {\em Feynman-{K}ac formulae: Genealogical and interacting particle
  systems with applications},
\newblock Springer, New York, 2004.

\bibitem{williams2006gaussian}
C.~K. Williams and C.~E. Rasmussen,
\newblock {\em Gaussian processes for machine learning},
\newblock MIT Press, 2006.

\bibitem{shahriari2016taking}
B.~Shahriari, K.~Swersky, Z.~Wang, R.~P. Adams, and N.~De~Freitas,
\newblock ``Taking the human out of the loop: A review of {B}ayesian
  optimization,''
\newblock {\em Proceedings of the IEEE}, vol. 104, no. 1, pp. 148--175, 2016.

\bibitem{kawaguchi2015bayesian}
K.~Kawaguchi, L.~P. Kaelbling, and T.~Lozano-P{\'e}rez,
\newblock ``Bayesian optimization with exponential convergence,''
\newblock in {\em Advances in neural information processing systems (NIPS)},
  2015, pp. 2809--2817.

\bibitem{liu2019robust}
B.~Liu,
\newblock ``Robust particle filtering via {B}ayesian nonparametric outlier
  modeling,''
\newblock in {\em Proc. of 22nd Int'l Conf. on Information Fusion (Fusion
  2019), in press}. IEEE, 2019.

\end{thebibliography}
\end{document}